\documentclass[letter,traditabstract,longauth]{aa}  
\usepackage{graphicx}
\usepackage{txfonts}
\usepackage{natbib} 
\bibpunct{(}{)}{;}{a}{}{,} 
\usepackage{longtable}
\begin{document}
\title{The far-infrared/submillimeter properties of galaxies\\
 located behind the Bullet cluster
\thanks{{\it Herschel} is an ESA space observatory with science instruments
provided by European-led Principal Investigator consortia and with important
participation from NASA.  Data presented in this paper were analyzed using
``The {\it Herschel} Interactive Processing Environment (HIPE),'' a joint
development by the {\it Herschel} Science Ground Segment Consortium, consisting
of ESA, the NASA {\it Herschel} Science Center, and the HIFI, PACS, and SPIRE
consortia.}}
   \author{M.~Rex\inst{\ref{inst1}} \and            
           T.~D.~Rawle\inst{\ref{inst1}} \and
           E.~Egami\inst{\ref{inst1}} \and
           P.~G.~P\'{e}rez-Gonz\'{a}lez\inst{\ref{inst14},\ref{inst1}} \and 
           M.~Zemcov\inst{\ref{inst3},\ref{inst4}} \and
           I.~Aretxaga\inst{\ref{instcc}} \and
           S.~M.~Chung\inst{\ref{instbb}} \and
           D.~Fadda\inst{\ref{inst9}} \and
           A.~H.~Gonzalez\inst{\ref{instbb}} \and
           D.~H.~Hughes\inst{\ref{instcc}} \and
           C.~Horellou\inst{\ref{instaa}}\and
           D.~Johansson\inst{\ref{instaa}}\and
           J.-P.~Kneib\inst{\ref{inst6}} \and
           J.~Richard\inst{\ref{inst15}} \and
           B.~Altieri\inst{\ref{inst2}} \and
           A.~K.~Fiedler\inst{\ref{inst1}} \and
           M.~J.~Pereira\inst{\ref{inst1}} \and
           G.~H.~Rieke\inst{\ref{inst1}} \and
           I.~Smail\inst{\ref{inst15}} \and
           I.~Valtchanov\inst{\ref{inst2}} \and
           A.~W.~Blain\inst{\ref{inst3}} \and 
           J.~J.~Bock\inst{\ref{inst3},\ref{inst4}}\and 
           F.~Boone\inst{\ref{inst5},\ref{inst7}} \and 
           C.~R.~Bridge\inst{\ref{inst3}}\and 
           B.~Clement\inst{\ref{inst6}} \and 
           F.~Combes\inst{\ref{inst7}} \and
           C.~D.~Dowell\inst{\ref{inst3},\ref{inst4}} \and
           M.~Dessauges-Zavadsky\inst{\ref{inst8}} \and 
           O.~Ilbert\inst{\ref{inst6}} \and
           R.~J.~Ivison\inst{\ref{inst10},\ref{inst11}} \and 
           M.~Jauzac\inst{\ref{inst6}}\and 
           D.~Lutz\inst{\ref{inst12}} \and
           A.~Omont\inst{\ref{inst13}} \and
           R.~Pell\'{o}\inst{\ref{inst5}} \and 
           G.~Rodighiero\inst{\ref{inst16}} \and
           D.~Schaerer\inst{\ref{inst8},\ref{inst5}} \and 
           G.~P.~Smith\inst{\ref{inst17}} \and 
           G.~L.~Walth\inst{\ref{inst1}} \and
           P.~van~der~Werf\inst{\ref{inst18}} \and
           M.~W.~Werner\inst{\ref{inst4}} \and
           J.~E.~Austermann\inst{\ref{instdd}} \and
           H.~Ezawa\inst{\ref{instee}} \and
           R.~Kawabe\inst{\ref{instee}} \and
           K.~Kohno\inst{\ref{instii},\ref{instjj}} \and
           T.~A.~Perera\inst{\ref{instgg}} \and
           K.~S.~Scott\inst{\ref{instff}} \and
           G.~W.~Wilson\inst{\ref{insthh}} \and
           M.~S.~Yun\inst{\ref{insthh}}                           }
 \institute{Steward Observatory, University of Arizona,
              933 N. Cherry Ave, Tucson, AZ 85721, USA;
              \email{mrex@as.arizona.edu}\label{inst1}
         \and
         Departamento de Astrof\'{\i}sica, Facultad de
         CC. F\'{\i}sicas, Universidad Complutense de Madrid, E-28040
         Madrid, Spain\label{inst14}
         \and
         California Institute of Technology, Pasadena, CA 91125,
         USA\label{inst3}
         \and
         Jet Propulsion Laboratory, Pasadena, CA 91109, USA\label{inst4}
         \and
         Instituto Nacional de Astrof\'isica \'Optica y Electr\'onica (INAOE),
         Luis Enrique Erro No.1, Tonantzintla, Puebla, C.P. 72840,
         Mexico\label{instcc}
         \and
         Department of Astronomy, University of Florida, Gainesville,
         FL 32611-2055, USA\label{instbb}
         \and
         NASA Herschel Science Center, California Institute of
         Technology, MS 100-22, Pasadena, CA 91125, USA\label{inst9}
         \and
         Onsala Space Observatory, Chalmers University of Technology,
         SE-439 92 Onsala, Sweden\label{instaa}
         \and
         Laboratoire d'Astrophysique de Marseille, CNRS -
         Universit\'{e} Aix-Marseille, 38 rue Fr\'{e}d\'{e}ric
         Joliot-Curie, 13388 Marseille Cedex 13, France\label{inst6}
         \and
         Institute for Computational Cosmology, Department of Physics,
         Durham University, South Road, Durham DH1 3LE, UK\label{inst15}
         \and
         Herschel Science Centre, ESAC, ESA, PO Box 78, Villanueva de
         la Ca\~nada, 28691 Madrid, Spain\label{inst2}
         \and
         Laboratoire d'Astrophysique de Toulouse-Tarbes,
         Universit\'{e} de Toulouse, CNRS, 14 Av. Edouard Belin, 31400
         Toulouse, France\label{inst5}
         \and
         Observatoire de Paris, LERMA, 61 Av. de l'Observatoire, 75014
         Paris, France\label{inst7}
         \and
         Geneva Observatory, University of Geneva, 51, Ch. des
         Maillettes, CH-1290 Versoix, Switzerland\label{inst8}
         \and
         UK Astronomy Technology Centre, Science and Technology
         Facilities Council, Royal Observatory, Blackford Hill,
         Edinburgh EH9 3HJ, UK\label{inst10}
         \and
         Institute for Astronomy, University of Edinburgh, Blackford
         Hill, Edinburgh EH9 3HJ, UK\label{inst11}
         \and
         Max-Planck-Institut f\"{u}r extraterrestrische Physik,
         Postfach 1312, 85741 Garching, Germany\label{inst12}
         \and
         Institut d'Astrophysique de Paris, CNRS and Universit\'{e}
         Pierre et Marie Curie, 98bis Boulevard Arago, F-75014 Paris,
         France\label{inst13}
         \and
         Department of Astronomy, University of Padova,
         Vicolo dell'Osservatorio 3, I-35122 Padova, Italy\label{inst16}
         \and
         School of Physics and Astronomy, University of Birmingham,
         Edgbaston, Birmingham, B15 2TT, UK\label{inst17}
         \and
         Sterrewacht Leiden, Leiden University, PO Box 9513, 2300 RA
         Leiden, the Netherlands\label{inst18}
	 \and 
         Center for Astrophysics and Space Astronomy, University of Colorado, 
	 Boulder, CO 80309, USA\label{instdd}
	 \and 
         Nobeyama Radio Observatory, National Astronomical Observatory of Japan,         Minamimaki, Minamisaku, Nagano 384-1305 Japan\label{instee}
         \and 
         Institute of Astronomy, University of Tokyo, 2-21-1 Osawa, Mitaka,
         Tokyo 181-0015, Japan\label{instii}
         \and
         Research Center for the Early Universe, School of Science,
         University of Tokyo,7-3-1 Hongo, Bunkyo, Tokyo 113-0033, Japan
         \label{instjj}
         \and
         Department of Physics, Illinois Wesleyan University,Bloomington, 
         IL 61702-2900, USA\label{instgg}
         \and
         Department of Physics \& Astronomy, University of Pennsylvania,
         209 South 33rd Street, Philadelphia, PA 19104, USA\label{instff}
         \and
         Department of Astronomy, University of Massachusetts, 
         710 North Pleasant Street, Amherst, MA 01003, USA\label{insthh}
}
\date{Received March 31, 2010; accepted May 18, 2010}
\abstract{The {\it Herschel} Lensing Survey  (HLS) takes advantage of
gravitational lensing by massive galaxy clusters to sample a population of
high-redshift galaxies which are too faint to be detected above the confusion
limit of current far-infrared/submillimeter telescopes. Measurements from
100--500\,$\mu$m bracket the peaks of the far-infrared spectral energy
distributions of these galaxies, characterizing their infrared luminosities
and star formation rates.  We introduce initial results from our science
demonstration phase observations, directed toward the Bullet cluster
(1E0657-56). By combining our observations with LABOCA 870\,$\mu$m and
AzTEC 1.1\,mm data we fully constrain the spectral energy distributions of 19
MIPS 24\,$\mu$m-selected galaxies which are located behind the cluster.  We
find that their colors are best fit using templates based on local galaxies
with systematically lower infrared luminosities.This suggests that our
sources are not like local ultra-luminous infrared galaxies in which
vigorous star formation is contained in a compact highly dust-obscured region.
Instead, they appear to be scaled up versions of lower luminosity local
galaxies with star formation occurring on larger physical scales.}
\keywords{Infrared: galaxies -- Submillimeter: galaxies -- Galaxies:
evolution -- Galaxies: high-redshift -- Galaxies: clusters: general --
Gravitational lensing:strong}
\maketitle
\section{Introduction}
\begin{figure*}[ht]
\centering
\includegraphics[width=5.0in]{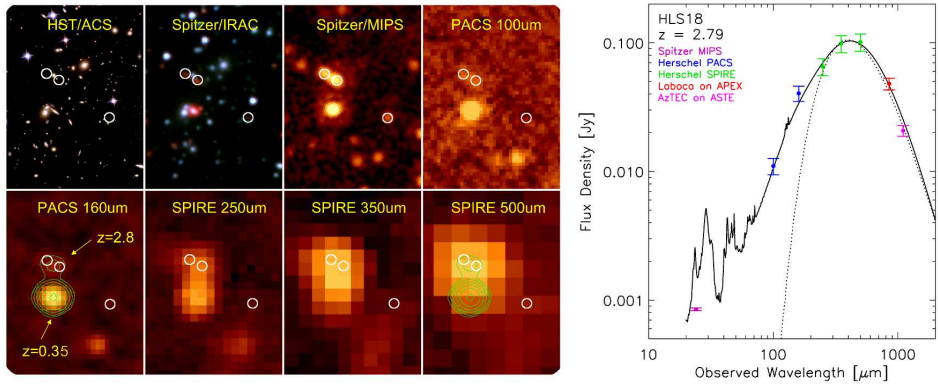}
\caption{The most highly magnified galaxy in our sample, HLS18,
is the same LIRG first detected by \citet{Wilson2008} with AzTEC at 1.1\,mm.
{\it Left panel}: 1.5$\arcmin \times $2$\arcmin$ thumbnails of the source at a
variety of wavelengths highlight the importance of multi-wavelength data to
guide the interpretation of the lower resolution maps.  IRAC data reveal three
lensed images of the galaxy indicated with white circles, although only
emission from the two brightest images is detected in the {\it Herschel} maps.
{\it Right panel}: \citet{Rieke2009} SED template fit to the source.  The
fluxes include the contribution from both of the brightest images of the
galaxy, indicated by the two left-most circles in the thumbnails.  The data
points have not been de-magnified. The 24\,$\mu$m flux is plotted for
reference, although it was not used in the fit.  The best-fit modified
blackbody is shown with a dotted line and is consistent with previous estimates
\citep{Wilson2008,Gonzalez2009,Rex2009}.
}
\label{lssrc}
\end{figure*}
Luminous (and ultra-luminous) infrared galaxies ((U)LIRGS) are a population of
massive star-forming galaxies which contain significant amounts of dust,
absorbing the ultraviolet emission from newly formed stars within them and
re-radiating this energy at far-infrared (FIR) wavelengths. The FIR emission of
these sources is therefore strongly correlated with their level of star
formation activity \citep[e.g.][]{Kennicutt1998}.  {\it {Spitzer}} MIPS
measurements from 24--160\,$\mu$m bracket the peak of the dust emission from
nearby (U)LIRGS.  Therefore, MIPS observations have been very successful in
characterizing the spectral energy distributions (SEDs) and star formation
rates (SFRs) of the local population \citep[e.g.][]{Rieke2009}.  Similar measurements
at FIR/submillimeter (submm) wavelengths are required to constrain the
SEDs of their higher redshift counterparts.  Together the PACS
\citep{Poglitsch2010} and SPIRE \citep{Griffin2010} instruments on-board {\it
Herschel} \citep{Pilbratt2010} provide sensitive measurements of this
population at five wavelengths from 100--500\,$\mu$m, sampling the rest-frame
peak of the FIR emission out to a redshift of $z = 4$.   

The depth of submm maps is ultimately limited by confusion
noise resulting from a high density of sources relative to the angular
resolution of the telescope.  Gravitational lensing by massive galaxy
clusters provides the only means to sample an abundant population of
intrinsically faint or high-redshift infrared (IR) galaxies which lie below 
this limit. The {\it Herschel} Lensing Survey (HLS) \citep{Egami2010} will
target $\sim$ 40 galaxy clusters with PACS and SPIRE to compile the first
significant submm catalog of these galaxies. In this letter we present
initial results from our science demonstration phase observations: 8
$\arcmin \times$ 8$\arcmin$ PACS maps and 17 $\arcmin \times$ 17$\arcmin$
SPIRE maps centered on the Bullet cluster (1E0657-56).  Details of these
observations are presented in \citet{Egami2010}.  This is a unique target
because of its strong lensing potential and the richness of existing ancillary
data.  We present the FIR properties of 19 MIPS 24\,$\mu$m-selected galaxies
which are located behind the Bullet cluster.  We compare the shapes of their
SEDs with those of local (U)LIRGS, and compare their measured IR
luminosities with predictions which extrapolate this quantity from the observed
MIPS 24\,$\mu$m flux. Subsequent uses of MIPS in the text refer to the MIPS
24\,$\mu$m band.
\section{Data and source catalog} The PACS and SPIRE observations of the 
Bullet cluster were reduced using the {\it Herschel} Interactive Processing 
Environment (HIPE).  Small deviations from the standard pipeline are discussed 
in \citet{Egami2010}. 
\subsection{Source selection} 
PACS and SPIRE data provide unprecedented sensitivities at FIR/submm
wavelengths.  Nonetheless, the precise identification of IR galaxies based
solely on {\it Herschel} maps is complicated due to relatively large beam sizes
leading to confusion noise in the SPIRE bands.  We have therefore constructed
an initial catalog with the positions of IR sources in the region based on
secure ($> 10$-$\sigma$) detections in the higher resolution {\it Spitzer} MIPS
24\,$\mu$m map of the field.  A 10-$\sigma$ 24\,$\mu$m threshold corresponds to
the 1-$\sigma$ error in the PACS 100\,$\mu$m map, our deepest {\it Herschel}
map of the field. We select the positions for our photometry from this MIPS
catalog in an effort to minimize the effect of Eddington bias which
artificially boosts the flux of sources selected from confused submillimeter
maps because of the steep underlying number counts.  A proper statistical
treatment of this effect will be presented in the more comprehensive analysis
of this field in preparation.

In order to identify a subset of these initial sources which lie behind the
Bullet cluster, we have compiled all of the spectroscopic redshift information
for galaxies in the field obtained thus far.  A description of these redshift
catalogs and our method of association with the MIPS positions is given in
\citet{Rawle2010}. A histogram of the redshifts reveals a large group of
Bullet cluster members at $z \sim 0.3$, as well as a smaller cluster of galaxies
located just beyond at $z \sim 0.35$ \citep{Rawle2010}.  We have
therefore chosen a lower limit of $z \ge 0.40$ to select background field
galaxies.  With these criteria we assemble a sub-catalog of 50 MIPS-selected
galaxies which are spectroscopically confirmed to be located behind the Bullet
cluster.  In many cases one SPIRE beam contains more than one MIPS galaxy.  In
these instances we are sometimes able to use the PACS resolution to identify
the dominant source of the SPIRE emission. If this is not possible we exclude
the source from our present analysis. Finally, we constrain our work to 
sources with $> 3$-$\sigma$ detections in at least two {\it Herschel} bands. 
We are thereby left with a sample of 15 significant background galaxies with 
spectroscopic redshifts. 

We also present the analysis of 4 additional galaxies selected because of
photometric redshift (photo-z) estimates suggesting they are at $z_{phot} >
1.5$, and which are detected in LABOCA 870\,$\mu$m \citep{Johansson2010} and
AzTEC 1.1\,mm maps of the field \citep{Wilson2008}.  We have calculated
photo-z's for these sources using two methods: one based on IRAC colors
\citep{PGonzalez2005} and the other on the FIR--mm colors \citep{Hughes2002,
Aretxaga2003}.  When the two estimates disagree we have chosen the value which
yields the lowest $\chi^2$ fit to our SED templates.  Including these galaxies
in our analysis allows us to take advantage of the substantial correlation
between SPIRE maps and longer wavelength submm/mm maps of the field.  These
results demonstrate the strength of combining such data sets to identify a
potentially higher redshift galaxy population.The positions of our selections,
along with corresponding MIPS and {\it Herschel} photometry can be found in
table~\ref{tab1} (included in the online supplementary material).
\subsection{Photometry}
\begin{figure*}[ht]
\centering
\includegraphics[width=\linewidth]{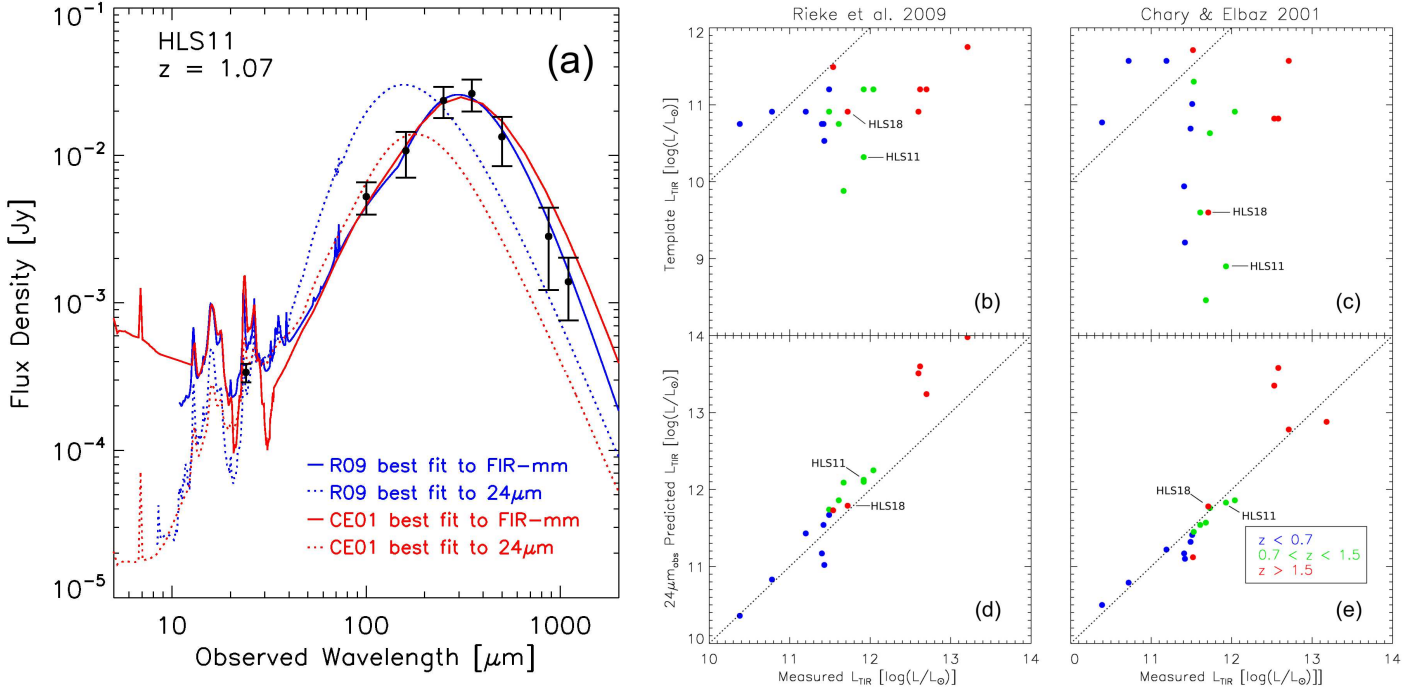}
\caption{{\it Left panel (a)}: SED template fits to HLS11, a redshift $z\sim1$
LIRG. The data points have been de-magnified by a factor of 1.4 based on our
lensing models of the foreground cluster.  Although both the R09 and the CE01
templates predict the IR luminosity based on observed 24\,$\mu$m emission to
within a factor of 2, the 24\,$\mu$m-predicted SEDs do not provide reasonable
fits to the FIR--mm data.
{\it Right top panels (b \& c)}: The IR luminosity
of the best-fit template as it corresponds to the local population is plotted
as a function of the actual IR luminosity of each galaxy.  The SEDs of our
sample are better fit by templates corresponding to systematically
lower luminosity local galaxies.  
{\it Right bottom panels (d \& e)}: The
24\,$\mu$m-predicted IR luminosity is plotted as a function of the actual IR
luminosity of each galaxy. Both template families predict reasonable values 
for the luminosities of the galaxies in our sample, except in the cases of the 
ULIRGS/Hyper-LIRGS at $z > 1.5$.  Note that HLS11 \& HLS18 (Fig.~\ref{lssrc})
fall on the trend line even though their SED shapes show significant deviation in (b) \& (c).  }
\label{panel}
\end{figure*}
\onltab{1}{
\begin{table*}
\caption{Flux densities measured from the MIPS and {\it Herschel} maps. The
500\,$\mu$m map was corrected for SZ contamination before extracting the
photometry. These measurements have not been corrected for the amplification
due to gravitational lensing. }
\label{tab1}
\centering
\begin{tabular}{llllllllll}
\hline
ID & RA & Dec & Flux$_{24}$ & Flux$_{100}$ & Flux$_{160}$ & Flux$_{250}$ & Flux$_{350}$ & Flux$_{500}$ \\
   &[deg] & [deg] & [mJy] & [mJy] & [mJy] & [mJy] & [mJy] & [mJy]\\
\hline
\hline
HLS01 &  104.72828 &  -55.88917 &      0.40$\pm$  0.01 &    $ \ldots$  &    $ \ldots$  &       21.1$\pm$    2.5 &       10.0$\pm$    3.5 &    $< 7.8$\tablefootmark{b} \\
HLS02 &  104.46419 &  -56.02209 &      0.15$\pm$  0.01 &    $ \ldots$  &    $ \ldots$  &       10.1$\pm$    2.4 &       16.2$\pm$    3.5 &       14.4$\pm$    4.0\\
HLS03 &  104.48783 &  -56.03070 &      0.39$\pm$  0.02 &    $ \ldots$  &    $ \ldots$  &        9.0$\pm$    2.4 &        9.5$\pm$    3.4 &    $< 7.8$\tablefootmark{b} \\
HLS04 &  104.45727 &  -55.91533 &      0.39$\pm$  0.01 &    $ \ldots$  &    $ \ldots$  &       18.6$\pm$    2.8 &       13.2$\pm$    3.6 &    $< 7.8$\tablefootmark{b} \\
HLS05 &  104.60260 &  -55.92013 &      0.71$\pm$  0.02 &       75.4$\pm$    1.3 &      164.4$\pm$    5.1 &      168.9$\pm$    6.2 &      120.0$\pm$    4.1 &       58.4$\pm$    4.4\\
HLS06 &  104.64300 &  -55.90990 &      0.40$\pm$  0.01 &        9.9$\pm$    1.2 &       28.4$\pm$    2.4 &       44.6$\pm$    2.8 &       39.0$\pm$    3.5 &       25.2$\pm$    4.0\\
HLS07 &  104.60545 &  -55.89051 &      0.29$\pm$  0.01 &        7.0$\pm$    1.3 &        7.8$\pm$    2.1 &        8.2$\pm$    2.8 &    $< 8.4$\tablefootmark{b}  &    $< 7.8$\tablefootmark{b} \\
HLS08 &  104.63628 &  -55.98351 &      0.135$\pm$  0.003 &        2.7$\pm$    1.1 &        4.6$\pm$    2.1 &    $< 6.0$\tablefootmark{b}  &    $< 8.4$\tablefootmark{b}   &    $< 7.8$\tablefootmark{b} \\
HLS09 &  104.55591 &  -55.87279 &      0.65$\pm$  0.02 &    $ \ldots$  &    $ \ldots$  &       23.2$\pm$    2.6 &       12.9$\pm$    3.5 &    $< 7.8$\tablefootmark{b} \\
HLS10 &  104.54350 &  -55.98005 &      0.53$\pm$  0.02 &       16.7$\pm$    1.4 &       22.6$\pm$    2.1 &       13.9$\pm$    2.4 &    $< 8.4$\tablefootmark{b}  &    $ < 7.8$\tablefootmark{b} \\
HLS11 &  104.56008 &  -55.95848 &      0.47$\pm$  0.01 &        7.4$\pm$    1.1 &       15.1$\pm$    2.1 &       33.1$\pm$    2.7 &       36.9$\pm$    3.5 &       18.8$\pm$    4.1\\
HLS12 &  104.62997 &  -55.94386 &      0.046$\pm$  0.003 &    $< 2.4$\tablefootmark{b}  &        4.3$\pm$    2.1 &        7.8$\pm$    2.5 &       11.4$\pm$    3.4 &    $ < 7.8$\tablefootmark{b} \\
HLS13 &  104.60567 &  -55.94490 &      0.28$\pm$  0.01 &        2.9$\pm$    1.1 &        5.5$\pm$    2.0 &       10.1$\pm$    2.5 &       19.6$\pm$    3.5 &    $ < 7.8$\tablefootmark{b} \\
HLS14 &  104.64689 &  -55.88658 &      0.136$\pm$  0.004 &        3.0$\pm$    1.2 &        4.4$\pm$    2.1 &    $< 6.0$\tablefootmark{b}  &    $< 8.4$\tablefootmark{b}  &    $< 7.8$\tablefootmark{b} \\
HLS15 &  104.73730 &  -55.88516 &      0.16$\pm$  0.02 &    $ \ldots$  &    $ \ldots$  &       11.1$\pm$    2.7 &       15.3$\pm$    3.8 &       11.7$\pm$    3.7\\
HLS16 &  104.58575 &  -55.93920 &      0.83$\pm$  0.02 &       25.3$\pm$    1.3 &       51.3$\pm$    2.5 &       47.7$\pm$    2.6 &       24.0$\pm$    3.6 &       10.2$\pm$    3.7\\
HLS17 &  104.64463 &  -56.00850 &      0.71$\pm$  0.02 &       24.7$\pm$    1.2 &       56.5$\pm$    3.4 &       49.2$\pm$    3.4 &       35.5$\pm$    3.7 &       17.9$\pm$    3.7\\
HLS18$_{A}$\tablefootmark{a} &  104.65471 &  -55.95193 &      0.49$\pm$  0.01 &        7.0$\pm$    1.2 &       24.5$\pm$    2.1 &       65.3$\pm$    2.8 &       98.6$\pm$    3.9 &      101.4$\pm$    4.0\\
HLS18$_{B}$\tablefootmark{a} &  104.65861 &  -55.95057 &        0.36$\pm$  0.01 &      3.8$\pm$  1.1   &       14.0$\pm$    2.5 &        $< 6.0$\tablefootmark{b}        &    $< 8.4$\tablefootmark{b}            &    $<7.8$\tablefootmark{b}\\

HLS19 &  104.59877 &  -55.87833 &      1.10$\pm$  0.03 &    $ \ldots$  &    $ \ldots$  &       40.9$\pm$    2.9 &       21.6$\pm$    4.0 &        9.4$\pm$    3.7\\

\hline
\end{tabular}
\begin{list}{}{}
\item[$^{\mathrm{a}}$] Multiple images of the highly lensed galaxy \citep{Gonzalez2009}.
\item[$^{\mathrm{b}}$] Upper limits correspond to statistical 3-$\sigma$ noise levels of each map.
\end{list}
\end{table*}
}
{\it Herschel} photometry for these galaxies is measured by 
simultaneously fitting the PSF at the positions of all of the 10-$\sigma$ MIPS
sources identified in our original catalog using routines included in the IRAF
package DAOPHOT. To ensure the maps are aligned to the same astrometry small
offsets are applied to the PACS data based on the results of stacking the
100\,$\mu$m map on the MIPS positions and to the SPIRE data based on the
results of stacking the 250\,$\mu$m map on the MIPS positions.  The 500\,$\mu$m
map is also corrected for contamination from the Sunyaev-Zel'dovich effect
based on fits to the data presented in \citet{Zemcov2010} before performing
photometry.
 
In cases where multiple MIPS galaxies fall within half the beam size at the
corresponding waveband, we approximate the position of the group by the average
of the counterparts weighted by their 24\,$\mu$m signal-to-noise.  We then take
an iterative approach; fitting at the positions of the brightest sources first,
removing them from the map, fitting at the positions of the next tier, and so
on until we have reached the 3-$\sigma$ noise level of the observation. This
method is similar to that outlined in \citet{PGonzalez2010}, although we force
the algorithm to fit the PSF at the MIPS positions rather than allowing for any re-centering.

Our analysis includes LABOCA 870\,$\mu$m and AzTEC 1.1\,mm photometry for our
sources when it is of $> 3$-$\sigma$ significance.  Four of our objects have
LABOCA counterparts within 8$\arcsec$ listed in the photometric catalog from
\citet{Johansson2010}.  In these cases we use the deboosted flux given in the
catalog for our analysis.  In order to obtain photometry for the remaining
sources, we measure the flux in an aperture of 40$\arcsec$ at the MIPS position
in order to get the total flux from the beam.  This method is not the same as
that used in \citet{Johansson2010}, although it gives consistent results for
the 17 sources presented therein.  In the case of the four sources chosen, in
part, due to bright long-wavelength emission, we identify clear associations
between the MIPS source and a significant, individual AzTEC source. In these
cases we use the deboosted flux from the AzTEC catalog.  For the other sources
the AzTEC fluxes are measured at the MIPS positions from a PSF-convolved map.
In the instances where there is a counterpart (within 8$\arcsec$) in the AzTEC
catalog, the difference between the flux at the MIPS position and the deboosted
point-source flux is $< 10\%$.
\section{Results}
\onlfig{3}{
\begin{figure*}[h]
\centering
\includegraphics[width=\linewidth]{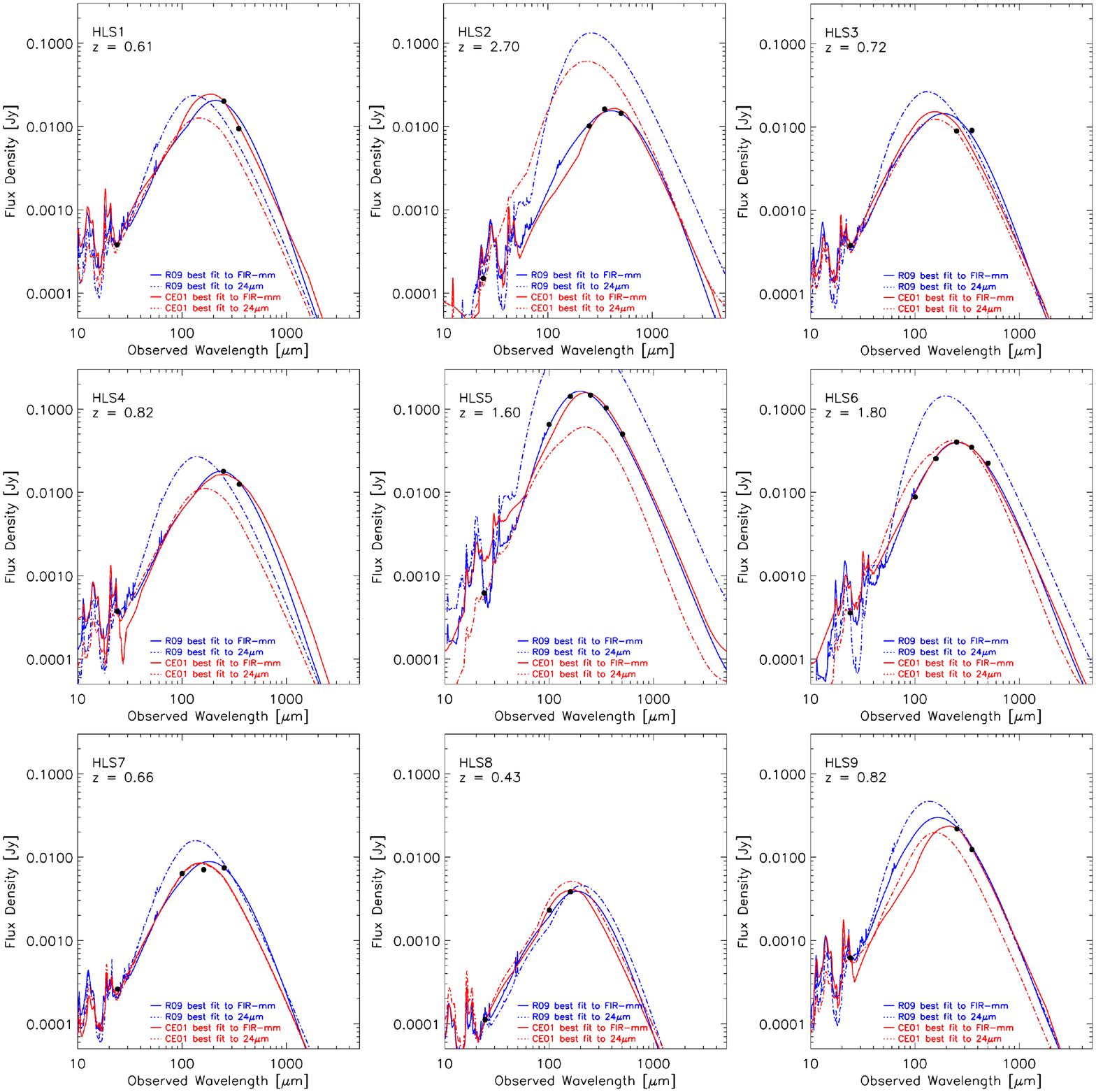}
\caption{SED template fits to HLS1--HLS9. The data points have been 
de-magnified according to our lensing models of the foreground cluster.  These
magnification factors can be found in table~\ref{tab2}. The solid lines show
the best-fit R09 template (in blue) and CE01 template (in red) to the FIR--mm
data, excluding the 24\,$\mu$m point.  LABOCA 870\,$\mu$m and AzTEC 1.1\,mm
data are used to constrain these fits when the detection is of $> 3$-$\sigma$
significance.  However, only the MIPS and {\it Herschel} data are shown in the
figure. The dotted lines show the respective fits based solely on the observed
24\,$\mu$m point.  Although both the R09 and the CE01 templates generally
predict reasonable values for the IR luminosity based on observed 24\,$\mu$m
emission, the 24\,$\mu$m-predicted SEDs typically do not provide reasonable
fits to the FIR--mm data.}
\label{seds1}
\end{figure*}
}
\onlfig{4}{
\begin{figure*}[h]
\centering
\includegraphics[width=\linewidth]{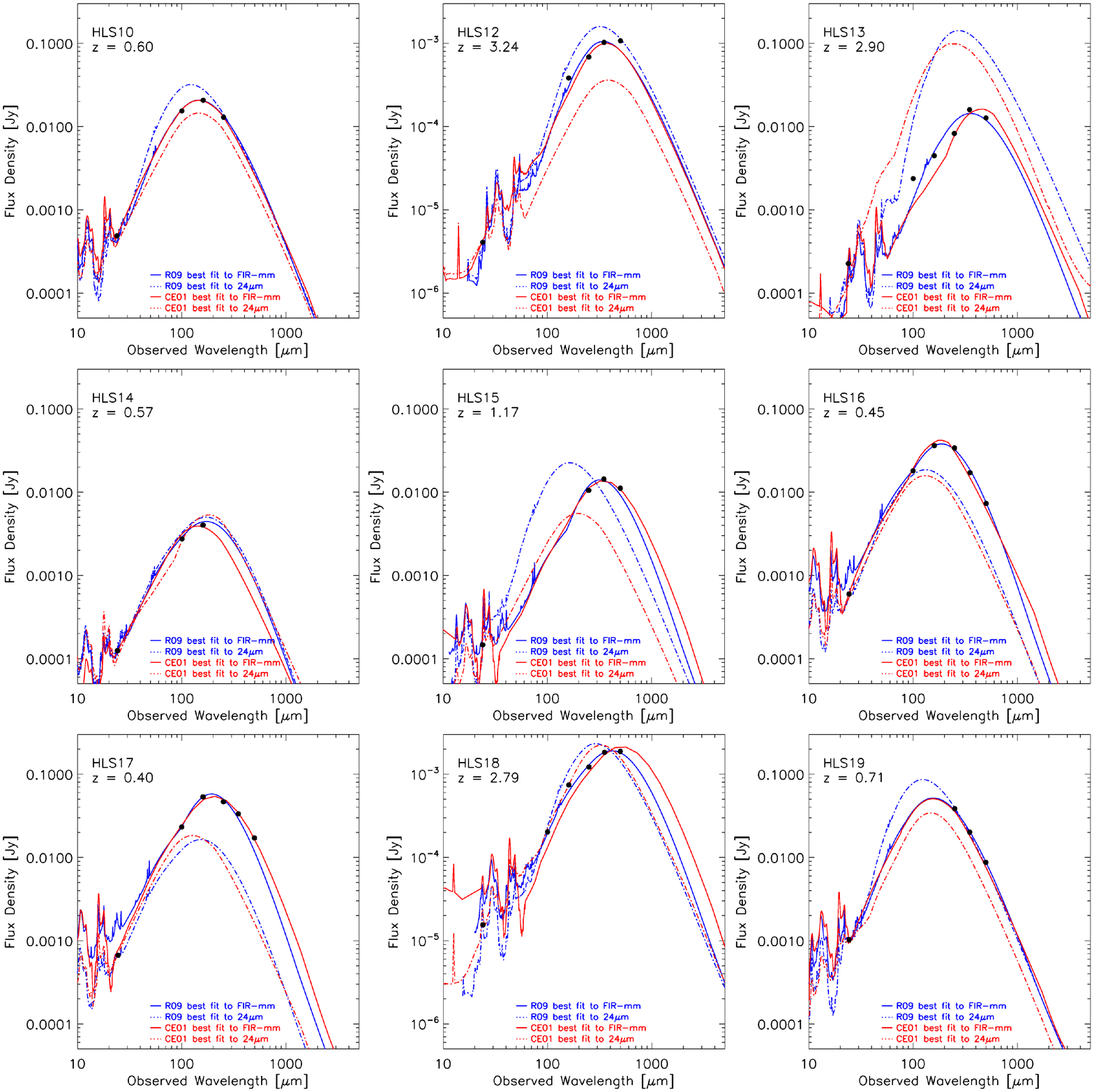}
\caption{SED template fits to HLS10--HLS19. The data points have been
de-magnified according to our lensing models of the foreground cluster.  These
magnification factors can be found in table~\ref{tab2}. The solid lines show
the best-fit R09 template (in blue) and CE01 template (in red) to the FIR--mm
data, excluding the 24\,$\mu$m point.  LABOCA 870\,$\mu$m and AzTEC 1.1\,mm
data are used to constrain these fits when the detection is of $> 3$-$\sigma$
significance.  However, only the MIPS and {\it Herschel} data are shown in the
figure. The dotted lines show the respective fits based solely on the observed
24\,$\mu$m point.  Although both the R09 and the CE01 templates generally
predict reasonable values for the IR luminosity based on observed 24\,$\mu$m
emission, the 24\,$\mu$m-predicted SEDs typically do not provide reasonable
fits to the FIR--mm data.}
\label{seds2}
\end{figure*}
}
Figure~\ref{lssrc} shows the most highly magnified galaxy in our sample, first
reported as a strongly lensed IRAC source \citep{Bradac2006,Gonzalez2009} and
independently found to be the brightest submm/mm source in the field
\citep{Wilson2008,Rex2009,Johansson2010}.  Multiple images of the galaxy are
indicated with white circles.  The right panel shows our SED fit to the sum of
the fluxes from the brightest of the two images.  SPIRE data indicate a
250\,$\mu$m flux density which is 30\% lower than that measured in the BLAST
data \citep{Rex2009}, although the measurements are consistent within the
uncertainties.  The {\it Spitzer} InfraRed Spectrograph (IRS) spectrum of this
source confirms that it is at redshift $z = 2.79$ \citep{Gonzalez2010},
consistent with published photo-z estimates similar to the ones used for the
four sources in our sample which lack spectroscopic redshift information.  At
that redshift, our model indicates a magnification of $\sim$ 50 due to
gravitational lensing by the foreground cluster.  This value is lower than
other estimates presented in the literature (e.g. \citet{Gonzalez2010} suggest
a magnification of $\sim$ 100).  We note that this value should be treated as a
lower limit since there is likely to be additional local lensing from nearby
objects.  We derive an observed IR luminosity of $2.8 \times 10^{13}$
L$_{\sun}$ based on the template SED fitting (see below for a detailed
discussion of this SED fitting procedure).  Adopting our estimated lower limit
for the magnification of the source indicates an intrinsic IR luminosity of
$\lesssim 5 \times 10^{11}$ L$_{\sun}$ which is consistent with calculations
previously reported in the literature
\citep{Wilson2008,Gonzalez2009,Rex2009,Gonzalez2010}, apart from the assumed
magnification factor. The detection of such an intrisically faint galaxy is not
possible without gravitational lensing, clearly demonstrating the power and
promise of our strategy.
\subsection{SED properties}
The galaxies in our sample span a redshift range of $0.40 < z < 3.24$.  Knowing
their redshifts, we can derive their apparent IR luminosities.  We use the
lensing models described in Paraficz et al. (in prep.) to correct
for the  magnification of each source due to gravitational lensing by the
foreground cluster and calculate the galaxies' intrinsic properties.  In our
discussion of IR properties, we adopt nomenclature similar to that in
\citet{Rieke2009}, defining the total IR luminosity (L$_{\mathrm {TIR}}$) as
the luminosity in the rest-frame wavelength range $\lambda= 5$--$1000\,\mu$m.
The redshifts, magnifications, and de-magnified IR luminosities of each source
are listed in table~\ref{tab2}. We assume a $\Lambda$CDM cosmology with
$\Omega_{m}=0.3$, $\Omega_{\Lambda}=0.7$, and $H_{0}=70$\,km s$^{-1}$Mpc$^{-1}$
for our calculations.

We use a $\chi^2$ minimization routine to fit the SED templates presented in
\citet{Rieke2009} (R09), as well as those presented in \citet{Chary2001}
(CE01). These templates are based on data from local galaxies and each is
designated by a luminosity class corresponding to the total infrared
luminosities of the local (U)LIRGS for which it provides the best fit.  The
left panel of Fig.~\ref{panel} shows an example of the template fits to a
galaxy in our sample.  The solid lines show the best-fit R09 template (in blue)
and CE01 template (in red) to the FIR--mm data, excluding the 24\,$\mu$m point.
The dotted lines show the respective fits to only the observed 24\,$\mu$m point.
The SED fits to the remaining galaxies in our sample are shown in 
Figs.~\ref{seds1} and \ref{seds2} (included in the online supplementary 
material).

In general we find that the templates provide good fits to the FIR--mm data in
our galaxy sample and that the IR luminosities derived using the best-fit
templates from both R09 and CE01 agree within a 1-$\sigma$ spread of 15\%.  We
also find that the galaxies in our sample are best fit by templates with
systematically lower luminosity classes.  The top right panels (b \& c) of
Fig.~\ref{panel} illustrate this trend.  The luminosity class of the best-fit
template to each galaxy is plotted as a function of the actual luminosity for
the R09 templates in (b) and the CE01 templates in (c). Although we find a
large scatter in the luminosity classes of best-fit SEDs, both template
families show the same general trend.  This is because the SEDs of our
galaxies peak at longer wavelengths than local galaxies of comparable
luminosities.  Therefore their spectral shapes more closely resemble those of
lower luminosity (U)LIRGS in the local universe.
The result supports evidence that high-redshift submm galaxies are
cooler than local galaxies with similar IR luminosities, suggesting the 
star formation within them is occurring on more extended physical scales
\citep[e.g.][]{Chapman2004}.

The discrepancy between the FIR SED shapes of our sources and those of local
counterparts with similar luminosities implies that a blind application of the
local template SED models might not accurately predict the $L_{\mathrm {TIR}}$
and hence the SFRs of higher redshift submm galaxies.  However, other studies
based on {\it Herschel} data show that the $L_{\mathrm {TIR}}$ predicted by
observed 24\,$\mu$m emission is in good agreement with that measured from the
FIR data for galaxies at $z < 1.5$ \citep{Elbaz2010}. The bottom right panels
of Fig.~\ref{panel} confirm this trend.  The IR luminosity derived from
24\,$\mu$m is plotted as a function of the actual IR luminosity for the R09
templates in (d) and the CE01 templates in (e). Although the R09 fits show a
slightly higher deviation, both template families predict reasonable values for
the luminosities of the galaxies in our sample, except in the cases of the
ULIRGS/Hyper-LIRGS at $z > 1.5$.  The left panel of Fig.~\ref{panel} shows a
closer inspection of the SED fits, revealing that although the predicted IR
luminosities are consistant with the measured values, the SED templates
selected based solely on observed 24\,$\mu$m emission typically peak at shorter
wavelengths, and do not provide good fits to the FIR--mm data. 
\section{Conclusion}
We have presented the first {\it Herschel} analysis of galaxies located behind
the Bullet cluster.  We find that their colors are best fit using templates
based on local galaxies with systematically lower IR luminosities.
This suggests that our sources are not like local ULIRGS in which vigorous star
formation is contained in a compact highly dust-obscured region. Instead, they
appear to be scaled up versions of lower luminosity local galaxies with star
formation occurring on larger physical scales.  A more comprehensive analysis
of the field in preparation will combine our full catalog of sources with
LABOCA and AzTEC data to compile a larger sample of these galaxies.  By
studying their relationship to local star-forming galaxies we can better
understand the processes that govern their evolution.
\begin{table}[h] 
\footnotesize
\caption{Source properties.} 
\label{tab2} 
\centering
\begin{tabular}{l l l l}   
\hline 
ID & z & Mag & R09 L$_{\mathrm{TIR}}$ [L$_{\sun}$] \\
\hline\hline 
HLS01 &       0.61 &       1.1 & 2.6e+11  \\
HLS02 &       \tablefootmark{a}{\bf {2.7$^{+.1}_{-.1}$}}, 4.1$^{+.7}_{-.6}$ &       1.00 & 4.0e+12  \\
HLS03 &       0.72 &       1.0 & 3.1e+11  \\
HLS04 &       0.82 &       1.0 & 4.1e+11  \\
HLS05 &       \tablefootmark{a}{\bf {1.6$^{+.1}_{-.1}$, 1.6$^{+.4}_{-.1}$}} &       1.16 & 1.6e+13  \\
HLS06 &       \tablefootmark{a}2.8$^{+.1}_{-.1}$,{\bf {1.8$^{+.5}_{-.4}$}} &       1.11 & 5.0e+12  \\
HLS07 &       0.66 &       1.1 & 1.6e+11  \\
HLS08 &       0.43 &       1.2 & 2.4e+10  \\
HLS08 &       0.82 &       1.1 & 8.3e+11  \\
HLS10 &       0.60 &       1.1 & 3.1e+11  \\
HLS11 &       1.07 &       1.4 & 8.3e+11  \\
HLS12 &       3.24 &      11.3 & 3.5e+11  \\
HLS13 &       \tablefootmark{a}{\bf {2.9$^{+.1}_{-.1}$}}, 3.1$^{+.7}_{-.3}$ &       1.23 & 4.2e+12  \\
HLS14 &       0.57 &       1.1 & 6.0e+10  \\
HLS15 &       1.17 &       1.1 & 4.7e+11  \\
HLS16 &       0.45 &       1.4 & 2.5e+11  \\
HLS17 &       0.40 &       1.1 & 2.7e+11  \\
HLS18 &       2.79 &       $>$54\tablefootmark{b} & 5.2e+11  \\
HLS19 &       0.71 &       1.1 & 1.1e+12  \\
 
\hline 
\end{tabular} 
\begin{list}{}{}
\item[$^{\mathrm{a}}$]IRAC photo-z, FIR-mm photo-z. Adopted value in bold.
\item[$^{\mathrm{b}}$]Lower limit due to unquantified local lensing by nearby objects.
\end{list}
\end{table}
\begin{acknowledgements}
We thank Ben Weiner and David Elbaz for their valuable comments and help with 
our computations.  This work is based in part on observations made with 
{\it Herschel}, a European Space Agency Cornerstone Mission with significant 
participation by NASA.  Support for this work was provided by NASA through an 
award issued by JPL/Caltech. 
\end{acknowledgements} 
\bibliographystyle{aa} 
\bibliography{refs}{}
\end{document}